# The iron record of asteroidal processes in carbonaceous chondrites

by

A. Garenne[1*], P. Beck[1,2], G. Montes-Hernandez[3], L. Bonal[1], E. Quirico[1], O. Proux[4], J.L. Hazemann[5].

[1]Univ. Grenoble Alpes, IPAG, F-38000 Grenoble, France

CNRS, IPAG, F-38000 Grenoble, France

[2]Institut Universitaire de France, Paris, France.

[3]Univ. Grenoble Alpes / CNRS-INSU, Institut des Sciences de la Terre (IsTERRE), Grenoble, France.

[4]Observatoire des Sciences de l'Univers de Grenoble (OSUG) CNRS UMS 832, 414 rue de la piscine 38400 Saint Martin d'Hères, France

[5]Institut Néel, 25 av. des Martyrs, 38042, Grenoble, France.

*Corresponding author

**Abstract**:

The valence of iron has been used in terrestrial studies to trace the hydrolysis of primary silicate rocks. Here, we use a similar approach to characterize the secondary processes namely thermal metamorphism and aqueous alteration of carbonaceous chondrites. X-ray Absorption Near-Edge Spectroscopy (XANES) at the Fe-Edge was performed on a series of 36 CM, 10 CR, 9 CV, and 2 CI chondrites. While previous studies have focused on the relative distribution of $Fe^0$ with respect to oxidized iron ($Fe_{ox}=Fe^{2+}+ Fe^{3+}$) or the iron distribution in some specific phases (e.g. Urey-Craig diagram (Urey and Craig 1953)) our measurements enable us to assess the fractions of iron in each of its three oxidation states: $Fe^0$, $Fe^{2+}$ and $Fe^{3+}$.

Among the 4 carbonaceous chondrites groups studied, a correlation between the Iron Oxidation Index (IOI = ($[Fe^0]+2[Fe^{2+}]+3[Fe^{3+}])/[Fe_{TOT}]$) and the hydrogen content is observed. However, within the CM group, for which a progressive alteration sequence has been defined, a conversion of $Fe^{3+}$ to $Fe^{2+}$ is observed with increasing degree of aqueous alteration. This reduction of iron can be explained by evolution in the mineralogy of the secondary phases. In the case of the few CM chondrites that experienced some thermal metamorphism, in addition to aqueous alteration, a redox memory of the aqueous alteration is present: a significant fraction of $Fe^{3+}$ is present, together with $Fe^{2+}$ and sometimes $Fe^0$.

From our dataset, the CR chondrites show a wider range of IOI from 1.5 to 2.5. In all considered CR chondrites, the three oxidation states of iron coexist. Even in the least altered CR chondrites, the fraction of $Fe^{3+}$ can be high (30 % for MET 00426). This observations confirms that oxidized iron iron has been integrated during fine-grained amorphous material formation in the matrix (Le Guillou et al. 2015; Le Guillou and Brearley 2014; Hopp and Vollmer 2018).
Last, the IOI of CV chondrites does not reflect the reduced/oxidized classification based on metal and magnetite proportion, but is strongly correlated with petrographic types. The valence of iron in CV chondrites therefore appears to be most closely correlated related to thermal history, rather than aqueous alteration even if these processes can occurs together (Krot et al. 2004; Brearley and Krot 2013).

1. Introduction

With the occurrence of three different oxidation states in planetary environments, the valence of iron has been used as a quantitative tracer of aqueous alteration processes on Earth (McCollom and Bach 2009; Malvoisin et al. 2012b; Malvoisin et al. 2012a). Corrosion of $Fe^{2+}$-bearing silicates in terrestrial hydrothermal systems leads to the production of secondary $Fe^{3+}$- bearing phases (some phyllosilicates, magnetite) coupled with the reduction of water and production of $H_2$ (Marcaillou et al. 2011).

Concerning chondritic meteorites, the Urey-Craig diagram, giving their relative iron contents in metal, sulfides, silicates and Fe-oxides which has been used to establish a classification (Weisberg et al. 2006; Krot et al. 2007). The initial budget of metallic iron in chondrites has been set by the compositions and relative abundances of the accreted components (type I and II chondrules, CAIs, metal/sulfides, matrix) and subsequently modified by parent body processes. The observed oxidation states in chondrites result from different processes in two distinct settings: protoplanetary disk and parent body (e.g. Lewis 1972; Rubin et al. 1988; Prinn and Fegley 1989; Zolensky et al. 1989; Bischoff 1998; Krot et al. 2000; Brearley 2006).

Primitive meteorites record the action of water (Brearley 2006) that led to significant alteration of the initially accreted phases. Although some alteration may have occurred in the solar nebula, the majority of the observed effects result from alteration on the asteroidal parent body (Bischoff 1998; Brearley 2014). In some carbonaceous chondrites, aqueous alteration can be pervasive and alteration signatures have been described in matrix, as well as in chondrule mesostasis (Brearley 2006). Meteorites from the CI and CM groups have a mineralogy that is dominated by secondary alteration products (Howard et al. 2009; King et al. 2015) and can contain up to 25 wt. % of different volatile elements in the case of Orgueil (Garenne et al. 2014). All CR chondrites have experienced aqueous alteration. The matrix appears to have been altered first, and chondrules show also aqueous alteration evidence observed as micrometric phyllosilicate rims for the least altered samples (Brearley 2006; Bonal et al. 2013; Harju et al. 2014; Le Guillou et al. 2015; Martinez and Brearley 2018). In CV chondrites, minerals resulting from low temperature metasomatism are less abundant but been unambiguously identified in some of them (e.g., Brearley 2006). These products differ between the oxidized and reduced CV subgroup (Howard et al. 2010).

The process of aqueous alteration leads to specific spectral signatures that can be searched for remotely across the small body population (e.g. small solar system bodies, such as asteroids, satellites, etc.). This includes a water-related feature around 3-µm (Rivkin et al. 2000; Beck et al. 2010), whose shape and intensity can be used to relate meteorites to asteroids. In addition, in the visible and near infrared (0.4-2.6μm), the observation of absorption features on C-type asteroids around 0.49, 0.70 and 0.90 μm are usually related to the presence of a significant amount of $Fe^{3+}$. These features are also observed in carbonaceous chondrites (Cloutis et al. 2011a; Cloutis et al. 2011b; Cloutis et al. 2012a ; Cloutis et al. 2012b). The valence of iron (and the mineral hosts) is the first order parameter that controls the visible-near infrared (VNIR) signature of chondrites.

In the present work, we determine the average iron oxidation index, calculated by $([Fe^0]+2[Fe^{2+}]+3[Fe^{3+}])/[Fe_{TOT}]$, in a series of primitive chondrites through synchrotron-based Fe-XANES spectroscopy . We estimate the abundance of the three iron oxidation states in bulk meteorite samples for wide range of petrologic typesand chondrite groups. We focus on carbonaceous chondrites for which aqueous alteration has been documented (CM, CV, CR and CI) in the literature.

## 2. Methods and samples
### 2.1 Samples

Samples studied here are from the CM, CV, CR and CI groups of carbonaceous chondrites (Table 1). The CIs, CRs and CMs cover a range of aqueous alteration with petrologic type ranging from 1.0 to 3.0 (Rubin et al. 2007; Howard et al. 2011). Among CM samples, some are identified as thermally metamorphosed (Nakamura 2005; Alexander et al. 2012) that caused modification of the secondary products and of the bulk iron oxidation state (Beck et al. 2011). CM chondrite falls and finds are compared to determine the potential oxidation induced by terrestrial residence. Reduced ($CV_{Red}$) and oxidized ($CV_{OxA}$ and $CV_{OxB}$) CV chondrites, that have experienced variable degrees of thermal metamorphism were selected as well.  Among our 9 CV chondrites, there are 3 $CV_{OxA}$ (Allende, Axtell, Mokoia), 3 $CV_{OxB}$ (Bali, Grosnaja, Kaba) and 3 $CV_{Red}$ (Efremovka, Leoville, Vigarano). Their petrologic types range from 3.1 to 3.6 (Bonal et al. 2006).

### 2.2 XANES spectroscopy

X-ray absorption spectroscopy is a powerful technique to study the physico-chemical properties of matter, including inter-atomic distance, coordination number, and electronic density. This method results from electronic transitions and photo-electron scattering with the neighbor atoms (single and multiple scatterings), allowing retrieval of structural and electronic information from the probed element.

X-ray Absorption Near-Edge Structure (XANES) spectra can be schematically described in three parts: (i) a pre-edge below edge-energy ($E_0$) corresponding to bound states transitions (mainly a 1s to 3d transition for the Fe K-edge), (ii) the absorption threshold (or edge jump) corresponding to transitions between the probed level to continuum states and/or to the last unoccupied bound levels (the edge crest, or the so-called white-line, corresponding to the 1s to 4p transition for the Fe K-edge) and (iii) the XANES part after the absorption edge corresponding to the scattering of the emitted photo-electron by the neighbor atoms.

The energy of the absorption edge will change when the core-level energies change. This absorption is specific for each element, allowing individual elements to be studied by selecting the appropriate experimental energy range parameter. X-ray absorption near-edge structure spectroscopy is very sensitive to oxidation state: as the oxidation state increases so does the absorption edge energy (Fig. 1).

## 2.3 Samples preparation and Fe-XANES measurements

Aliquots of 20 mg from a larger samples of meteorite powder were typically mixed with 30 mg of boron nitride (BN) and pressed in order to obtain a compact pellet with a diameter of 5 mm and a thickness of 2 mm. Each mixture of meteorite with BN was manually ground for 30 minutes in an agate mortar before pressing in order to ensure a homogeneous mixing. Each sample was analyzed for Fe-K-edge XANES at CRG-FAME beamline at the European Synchrotron Radiation Facility (Proux et al. 2005; Proux et al. 2006). The liquid nitrogen-cooled two-crystal monochromator is equipped with Si(220) crystals. Energy resolution is close to the intrinsic resolution of these crystals, i.e. 0.365 eV at the Fe K-edge. Spectra were measured in transmission mode between 7004 and 7385 eV with a typical energy step of 0.2 eV around the pre-edge and 2 eV in the further part (before 7105 eV and after 7150 eV). Energy calibration was achieved by setting the maximum of the first derivative of a Fe metal foil K-edge spectrum at 7112.0 eV.

In order to assess the homogeneity of each pellet, a scan in transmission above the iron absorption edge was performed across the sample, and three XANES spectra were measured at different locations. For each sample, the three XANES spectra were compared and averaged to have a good estimation of the iron oxidation state in the meteorite sample. The raw spectra were normalized using two polynomial curves, for the pre-edge region (linear curve) and the post-edge region (polynomial order: 2) while the edge jump was scaled by a factor to give a jump of 1. These normalizations were performed with the software ATHENA from the IFEFFIT package (Ravel and Newville 2005).

**2.4 Calculation of Fe oxidation state**

Several processing methods were tested on our spectral set to determine the optimum technique to derive Fe-oxidation state of the samples. In the literature, the pre-edge feature is often used: the combination of its intensity (for a normalized spectra) with its centroid position is used to infer the average valence and the site geometry for iron oxidized atoms (e.g. Wilke et al. 2001). However since significant amounts of $Fe^0$ are expected in our CR and CV samples, this approach was rejected.

A series of alternative methods to quantify the Fe-oxidation factor were tested, based on (i) the edge position from the derivative of the spectra, (ii) a linear combination of standards on the main part of XANES spectra (from -20 eV to +30 eV compared to $E_0$), (iii) a linear combination of standards of the pre-edge feature (from -15 eV to -5 eV compared to $E_0$). The pre-edge, expected to be less sensitive to surrounding atoms than the whole spectra, is more sensitive to the valence of iron. The method based on a linear combination of standards for the pre-edge feature was used (Beck et al. 2012). Therefore the average valence calculated for our samples is the weighted arithmetic mean of the used standards.

Our suite of standards includes iron metal ($Fe^0$), San Carlos olivine ($Fe^{2+}$), greenalite ($Fe^{2+}$), fayalite ($Fe^{2+}$), troilite ($Fe^{2+}$), pyrite ($Fe^{2+}$), cronstedtite ($Fe^{2+}+Fe^{3+}$), magnetite (1/3 $Fe^{2+}$+ 2/3 $Fe^{3+}$), ferrihydrite ($Fe^{3+}$) and hematite ($Fe^{3+}$). For a full description of these standards see Beck et al.( 2012).

**3. Results**

**3.1 Fe-XANES spectra of bulk carbonaceous chondrites**

Representative Fe-XANES spectra of CI, CM, CR and CV chondrites are compared in Figure 1a. They reveal the diversity of iron oxidation and Fe-bearing phases in these four carbonaceous chondrite groups. The Fe-XANES spectra of standards for $Fe^0$, $Fe^{2+}$ and $Fe^{3+}$ are in Figure 1b for comparison. An increase in energy of the edge position (around intensity of 0.9) can be used as a first way to assess the average oxidation state of these samples. In contrast to the Urey-Craig diagram (Urey and Craig 1953), this study combines the information of the three oxidation states in bulk sample. The value here is then strongly dependent on the complexity of the mineralogy and the co-existing phases, possibly evolving at the same time but in different ways than the oxidation state.

As expected, the CI chondrite (Orgueil) XANES spectra reveal the presence of oxidized Fe-bearing minerals. The CV chondrite (Axtell) appears to be the most reduced, while the CM (MET 01070) and the CR (Renazzo) appear to have similar edge positions. This position is strongly dependent on the metal contained in CRs chondrites. The energy position of the maxima of these spectra appears to increase with the abundance of metal, similarly to the behavior observed for the edge position.

The pre-edge region is valuable in deciphering the average valence of iron (Wilke et al. 2001). The spectra of the CR chondrites are distinct from the other spectra by showing an enhanced pre-edge intensity (Figure 1a), which is typical of the presence of some amount of metallic iron ($Fe^0$) (Figure1b). The pre-edge region of Orgueil (CI) also shows an important pre-edge intensity, but the position (around 7114 eV) and shape of the feature is more consistent with $Fe^{3+}$ rather than metallic iron (Figure 1b). The pre-edge features of the CM and the CV both show two peaks in this spectral region, but with different relative intensities (see inset in Fig. 1a).

Spectra obtained from CR chondrites reveal an important diversity in the intensity of the pre-edge features (Fig. 2). This feature is related to variable amounts of metal (Weisberg et al. 1993), the largest amount being in MET 00426. The spectrum of the CR1 GRO 95577 shows a much smaller pre-edge intensity with a maximum around 7114 eV, typical of the presence of $Fe^{3+}$. The pre-edge feature of Renazzo is intermediate between GRO 95577 and MET 00426.

Spectral variability among CMs (Fig. 3) is less obvious than for CRs. This is easily explained by the absence of significant amounts of metal (Rubin et al. 2007; Scott and Krot 2014). Still some variation is seen in terms of the edge position, the energy at the absorption maxima, and the pre-edge spectral shape. QUE 97990 (CM 2.6) seems to have the most oxidized mineralogy, MET 01070 (CM 2.0) appears intermediate, and finally PCA 02012 (heated CM) seems to have the most reduced Fe-bearing mineralogy (Fig. 3).

**3.2 The average iron oxidation index**

From the linear decomposition analysis of the pre-edge region, the proportion of each iron oxidation state, an average iron oxidation index and the $Fe^0/Fe_{tot}$ to ($Fe^{3+}/Fe^{2+}$) ratio were estimated (Table 1). The average "Iron Oxidation Index" (IOI) was calculated by the following equation:

$$IOI = ([Fe^0]+2[Fe^{2+}]+3[Fe^{3+}])/[Fe_{TOT}]$$

This parameter is strictly equivalent to the bulk valence used for instance by Sutton et al. (2017). However, we prefer to refer to it as "iron oxidation index" since a non-integer valence is not consistent with chemical rules.

The average IOI varies between 1.60 (CR chondrite GRA 06100) and 2.61 (CI chondrite Orgueil) among our samples and increases along the sequence CR < CV < CM < CI. This pattern is consistent with the higher abundance of metal in CR chondrites (8 vol. %, Krot et al. 2007) than in CI, CM and CV chondrites. Moreover, extensively aqueously altered CI and CM chondrites have the highest valence of iron.

The bulk valence derived by Sutton et al. (2017) with a similar method are in good agreement with the present IOI. Indeed, for Orgueil we obtained an IOI of 2.61 in comparison to 2.77 (Sutton et al. 2017), and for Renazzo 1.85 in comparison to 1.91 (Sutton et al. 2017). The IOI we derived for the six CM chondrites analysed in both studies appear to be lower with an average of 2.36 against 2.49 in Sutton et al. (2017). Note that the fitting methods differ between these two studies and our standard-based fits are performed in the pre-edge region only. Indeed the higher energy part is sensitive to interaction with neighboring atoms, and thus less related to the iron valency. The values obtained for the eight samples common to the two studies show a good correlation with an $R^2$ value of 0.90 (slope=0.99).

### 3.3 $Fe^0/Fe_{tot}$ to ($Fe^{3+}/Fe^{2+}$)

The matrix, being made of fine-grained material, is the petrological component most sensitive to aqueous alteration (e.g. McSween 1979; Browning et al. 1996). The abundance of matrix in each bulk sample may then affect the IOI. When comparing the metal proportion against the fraction of oxidized iron (Figure 4.b), a clear separation between the different chondrite groups appears. The CMs plot along the $Fe^{3+}/Fe^{2+}$ axis with ratios varying between 0.3 and 1.4, the CRs on the top of the diagram along $Fe^0/Fe_{tot}$ with ratios varying between 16 and 22% and the CVs are on the lower part of the diagram, closest to 0 for both ratios. CIs plot on the $Fe^{3+}/Fe^{2+}$ axis with a ratio of 1.47 and 0.70 for Orgueil and Ivuna, respectively. In contrast to the IOI, this diagram allows a clear distinction between CV and CR chondrites. This distinction results from a combination of higher proportions of $Fe^{3+}$ and metal in CR chondrites than in CV chondrites. The $Fe^0/Fe_{tot}$ ratio is also systematically higher in heated CMs than in non-heated CMs (metal was detected in five heated chondrites) (Fig 4.b)

The variability in IOI between and within chondrite groups is discussed in the following section

### 4 Discussion

**4.1 Group systematics, redox record and $H_2$ production in unheated CI, CM and CR chondrites.**

The interaction between an unaltered carbonaceous chondrite and water is expected to enrich it in $Fe^{3+}$ due to the corrosion of metallic iron. The dissolution of $Fe^{2+}$-bearing anhydrous silicates will be followed by the precipitation of $Fe^{2+}$- and $Fe^{3+}$-bearing phases (e.g. (oxy)hydroxide and phyllosilicates). While Fe is oxidized, water molecules can be reduced. This reaction is a possible source of molecular hydrogen ($H_2$), and can be associated with hydrogen isotopic fractionation. Fingerprints of this process have been possibly identified in ordinary chondrites in the form of D/H enrichment of the IOM (Insoluble Organic Matter) (Eiler and Kitchen 2004; Alexander et al. 2007; Sutton et al. 2017).

The CI chondrites that have experienced the most intense aqueous alteration tend to have the most oxidized iron mineralogy (Fig. 4). In particular CI and CM chondrites, which are all classified as types 1 or 2 (Rubin et al. 2007; Scott and Krot 2014), have the highest average IOI (Fig 4.a). The most oxidized chondrite is Orgueil. However the two CIs, Orgueil and Ivuna have very different $Fe^{3+}$ contents (61% and 41%, respectively) (Beck et al., 2012). The iron mineralogy of these meteorites is quite different: Orgueil contains some ferrihydrite and Ivuna does not. This difference could have partly resulted from post-terrestrial alteration (Gounelle and Zolensky 2001; Airieau et al. 2001; wiik 1956 ). Beck et al. (2012) found that 80% of the iron in Orgueil is present as oxides which is not the case of Ivuna which has a different mineralogy (see also Beck et al. 2014).

While several proxies and aqueous alteration schemes have been defined for carbonaceous chondrites, a convenient quantitative tracer of aqueous alteration extent is the amount of $H_2O$ and –OH trapped in minerals (Howard et al. 2009; Howard et al. 2011; Alexander et al. 2013). IOI seems to be roughly correlated to the bulk H abundance in CI, CR and CM chondrites as measured by Alexander et al. (2012) (Figure 5): the more aqueously altered the sample is, the more oxidized the iron is and CI, CM and CR chondrites plot apart. An important spread is observed within CMs and for these samples the IOI appears to be anticorrelated with aqueous alteration. The CV chondrites are not shown on Figure 5 since H data are only available for one sample (Kaba, that falls outside of this trend). While a general increase of H -content with the valence of iron is expected, the presence of a single trend for CR, CI and CM (average) is more surprising. The slope on this diagram (Fig. 5) is related to the redox reaction leading to the production of $Fe^{3+}$ and thus depends on the IOI of the sample prior to alteration and the nature of Fe-bearing secondary products. Since the secondary alteration products and their proportions are different among CI, CM and CR chondrites (Scott and Krot 2014; Bonal et al. 2013; Abreu 2016), the presence of a single trend is unexpected (i.e. abundant magnetite in CIs, but rare in CMs, amorphous secondary phases in CR, higher serpentine and saponite proportion in CM and CI than CR). We cannot exclude a statistical effect due the few CI and CR chondrites in our study compared to CM chondrites. A more complete study of CIs and CRs is necessary to determine wether or not this is a single trend.

During aqueous alteration, $H_2$ production might occur on the asteroidal parent body. The production of $H_2$ can be a first step before the synthesis of complex organics through Fischer-Tropsch Type reactions. This $H_2$ production can also impact the evolution of asteroids since pressure build-up could lead to partial or total disruption even if a fast $H_2$ degassing is possible (Guo and Eiler 2007; Wilson 1999).

**4.2 The reduction of Fe during aqueous alteration of CM chondrites**

CM chondrites have all experienced aqueous alteration, but to various extents (e.g. Zolensky et al. 1997; Rubin et al. 2007). While it is not proven that all CMs come from a single parent body (many large CM-like asteroids are present in the main-belt, Fornasier et al. 2014; Vernazza et al. 2016), these variable degrees of aqueous alteration are usually seen as a progression, from midly altered (unaltered mafic silicate phenocrysts in chondrules, rare metal grains preserved) to heavily altered (fully altered chondrules) (Rubin et al. 2007).

A majority of the considered CMs are Antarctic finds. The Fe valence could have been modified through terrestrial weathering (Lee and Bland 2004; Bland 2006). The average valence of iron is almost identical for CM falls (0.38, n=8) and CM finds (0.36, n=16), suggesting that terrestrial residence had a minimal impact on the Fe valence. However, we could not exclude a possible minor impact, hidden by the spread and variation of Fe-valence of these samples due to different degree of aqueous alteration.

Focusing on the CMs, the average Fe oxidation factor is compared to the hydrogen content (Alexander et al. 2012) (Fig. 5) and to the intensity of the 3-µm band (Fig. 6) that is related to –OH and $H_2O$. In both cases, the average IOI appears to decrease with increasing degree of aqueous alteration or more precisely, water content. This pattern shows that iron is reduced as aqueous alteration progresses in CM chondrites. The decrease of the average IOI with the extent of aqueous alteration in CM chondrites (Fig. 6) is at first unexpected. A similar effect has been observed on CR matrix chondrites by Le Guillou et al. (2015). As stated earlier, in the case of the aqueous alteration of terrestrial ultramafic rocks, the average $Fe^{3+}$ content increases with the degree of serpentinization (Marcaillou et al. 2011). However, CM chondrites are different from terrestrial ultramafic rocks that their protolith contained iron metal, and quite unusual alteration products were formed, such as tochilinites (an hydroxysulfide consisting of interlayer mackinawite and brucite layers), as well as cronstedtite.

To understand the reduction of iron, it is important to consider the chemistry of iron–bearing alteration products. Cronstedtite is a Fe-rich serpentine, which contains a significant amount of $Fe^{3+}$ in tetrahedral sites (Buseck and Hua 1993; Beck et al. 2010). Cronstedtite is abundant in the least-altered CM chondrites (Howard et al. 2011), but as aqueous alteration progresses, crondstedtite is progressively replaced by Mg-rich serpentine (Browning et al. 1996) as revealed by transmission electron microscopy observations. Mg-rich serpentine may contain $Fe^{3+}$, but not to the level of cronstedtite (Burns and Fisher 1994). This mineralogical evolution has been explained by the following sequence (Howard et al. 2011; Le Guillou et al. 2015): i) formation of cronstedtite from alteration of amorphous silicate and metal; ii) alteration of chondrules (Mg- and Si-rich) and increase of silica activity; iii) dissolution of cronstedtite and precipitation of Mg-rich serpentine and sulfides, iv) iron exchange from cronstedtite to serpentine in chondrules and substitution of iron by magnesium in cronstedtite.

The reduction of iron during aqueous alteration seems to be in agreement with this aqueous alteration sequence. In the least altered CM chondrites a significant fraction of $Fe^{3+}$ can be accommodated in cronstedtite which is much less abundant in the most-altered CMs. In the latter case, the mineralogy is dominated by Mg-serpentine (which can host only a limited amount of $Fe^{3+}$), and $Fe^{2+}$-bearing sulphides (Howard et al. 2009; Howard et al. 2011; Howard et al 2016).

The fact that on average the IOI decreases with aqueous alteration requires that iron should be reduced at the expense of other elements being oxidized (if the evolution occurs in closed-system). There are many possible reducing agents, the first one being dissolved $H_2$. The production of $H_2$ was likely significant in the first stages of aqueous alteration (from corrosion of metals) and an increase in hydrogen fugacity will destabilize cronstedtite toward $Fe^{2+}$-bearing serpentine greenalite (Zolotov 2014). Locally produced $H_2$ could therefore act as a reducing agent, under the assumption that it does not escape. A system under pressure would have facilitated the buildup of an $H_2$ overpressure. However, closed system alteration is an approximation and does not take into account partial escape of highly diffusing gases like $H_2$ (Krot et al. 2015).

An alternative scenario to explain the reduction of iron would be interaction with nebular $H_2$. But this would require not only that aqueous alteration took place before nebular gas dissipation (which contradicts Mn/Cr chronology of alteration products, Fujiya et al. 2012), but also that some mechanism would permit interaction of the subsurface of the asteroid and the surface. Therefore a reaction with nebular $H_2$ is unlikely. An alternative scenario is internal oxidation of different compounds on the asteroid surface reduce iron. It could be elemental sulfur, sulfides or oxysulfide into sulfates, though the sulfate amounts are likely low in carbonaceous chondrites (estimated between 0.4 and 1 wt. %; Burgess et al. 1991). Note that the sulfide abundance appears to decrease with extent of aqueous alteration (Burgess et al. 1991; Orthous-Daunay et al. 2010; Brearley 2011; Singerling and Brearley 2016). Finally, other compounds such as organic matter or dissolved CO could have acted as possible iron reducing agents in CMs.

The hydrolysis of iron metal can produce significant isotopic fractionation if the system is opened to $H_2$ escape (Alexander et al. 2010). In Figure 7, the oxidation factor of iron is compared to the bulk D/H measured in carbonaceous chondrites. This bulk value is the average of all H-bearing phases in carbonaceous chondrites, i.e. $H_2O$ and -OH bearing minerals, soluble and refractory organics. In the case of CM chondrites, a positive correlation is found between bulk D/H and iron oxidation index. The simplest explanation is that this diagram reflects mixing between phyllosilicates and organics, as discussed in Alexander et al. (2012). This would mean that the D/H of water in CR and CI chondrites is distinct from the water in CM chondrites (Alexander et al. 2012). An alternative would be that this trend is related to a chemical process, which changes both the isotopic composition of water and the IOI. A possible process is the reaction between water and nebular $H_2$ (with a low D/H), but this seems challenging given the reasons discussed above.

**4.3 Heated CM and the redox memory**

Among the suite of CM chondrites studied here, some are described as heated (Table 1). The existence of such samples was initially reported following the Japanese meteorite recovery campaigns in Antarctic (Akai 1992). While the petrology points toward affinities with CM chondrites, specific mineralogical changes demonstrate that these meteorites have undergone heating, based on mineralogy, texture, oxygen isotopes or organic matter (Nakamura 2005). The most likely source of thermal metamorphism is impact heating (Nakamura 2005). When considering the IOI, the majority of heated CM samples are indistinguishable from mainstream CM chondrites. However, when considering $Fe^0/Fe_{tot}$, it appears that half of heated CMs contain metallic irons in contrast to unheated CMs (Table 1). Two (paired) meteorites are quite distinct by their high metal abundance, PCA 02010 and PCA 02012. PCA 02010 and PCA 02012 appear to be the most reduced CM chondrites with the highest amount of $Fe^0$ (7-9 at. % of Fe). These samples may have experienced more intense heating than the other heated CMs studied here (Nakato et al. 2013; Beck et al. 2014) after they were originally hydrated (Nakamura 2005). It is difficult to determine whether the metal in heated CMs is a relic of unaltered initial iron or a product of metal formation after heating (Nozaki et al 2006). In the case of CV chondrites like Allende, the fluid-assisted thermal metamorphism seems to have oxidized the initial metal to $Fe^{2+}$ (Krot et al. 2004; Lee et al. 1996). With the exception of the proportion of metal, the $Fe^{2+}$ and $Fe^{3+}$ budgets of heated CMs are similar to those in unheated CMs. It seems that for low levels of thermal alteration, the oxidized nature of iron is preserved, and an "iron oxidation state memory" of the aqueous process is present.

**4.4 CV chondrites, discrepancy between iron valence and the red-ox dichotomy classification**

All CV chondrites are classified as type 3, but they have experienced various degrees of aqueous alteration and thermal metamorphism. They have been classified into two subgroups, so called "oxidized" ($CV_{Ox}$) and "reduced" ($CV_{Red}$), based on the relative abundance of metal and magnetite, as well as the Ni content of metal and sulfides (Mcsween 1977). Oxidized CV chondrites were later subdivided into $CV_{OxA}$ (Allende subgroup) and $CV_{OxB}$ (Bali subgroup) (Weisberg et al. 1997), where $CV_{OxA}$ typically have higher metal content and lower matrix abundance than $CV_{OxB}$

Among the studied samples, Efremovka, a reduced CV, contains the highest amount of metal, and no $Fe^{3+}$, while Kaba, classified as a $CV_{OxB}$, does indeed contain a high abundance of $Fe^{3+}$, the highest among our series of CVs (Table 1). However, considering the Fe oxidation factor, many samples do not follow the general classification scheme of CV oxidized and reduced (Table 1). Indeed, Vigarano ($CV_{Red}$) seems to have a lower metal abundance and higher $Fe^{3+}$ content than Bali ($CV_{OxB}$). Also, from the Fe oxidation factor perspective, Allende ($CV_{OxA}$) and Grosnaja ($CV_{OxB}$) seem to be similar. The iron oxidation in CV chondrites seems more complex than other carbonaceous chondrites. One explanation is that these materials have been altered by both aqueous and thermal metamorphism. The intensities of these geological processes are not the only parameters to take account, their chronological order too (Lee et al. 1996).

The oxidized/reduced classification is based on magnetite and metal contents only: other iron phases are not taken into account. These discrepancies have been already observed (Howard et al. 2010) and could result from iron being present in other mineral phases such as phyllosilicates. Also some of the observed variability falls within our analytical uncertainty on the retrieval of the various proportions of iron species (+/-10 %). The Fe oxidation state appears to be controlled by the thermal metamorphism and petrologic type proposed by Bonal et al. (2006).

**4.5 The oxidized matrix of CR chondrites**

The presence of a high abundance of presolar grains and the preservation of amorphous material in the matrix shows that some CR chondrites are amongst the most primitive meteorites in our collections (Floss and Stadermann 2009; Abreu and Brearley 2010). All of the CR chondrites in this study are classified as type 2 with the exception of GRO 95577, classified as a CR1 (Weisberg and Huber 2007; Harju et al. 2014). The CR1 sample is clearly distinct from the other studied CRs in terms of bulk H vs average iron oxidation index and falls within the array defined by CM chondrites (Fig. 5). This is not unexpected given the fact that GRO 95577 is the only CR sample that experienced extensive aqueous alteration..

The analysis of Fe-XANES spectra from our sample suite (excluding GRO 95577) shows that each CR contains iron metal, typically around 20 at. % of total iron. RBT 04133 is significantly different with a much smaller fraction of metal (5.6 at. %). This sample was previously described as anomalous by Davidson et al. (2009) from its oxygen isotope composition, with closer affinities to CV-CK than CR chondrites. It is metamorphosed (type 3.6 proposed by Bonal et al. 2016) and was proposed to be reclassified as $CV_{red}$ (Davidson et al. 2014). With the exception of GRA 06100, all CR samples have a significant fraction of iron present as $Fe^{3+}$. Note that GRA 06100 has been previously identified as a metamorphosed CR given its mid-infrared spectrum (Abreu and Bullock 2013; Briani et al. 2013; Beck et al. 2014).

In contrast to CVs, the CR chondrites have an important proportion of $Fe^{3+}$, closer to CMs, probably resulting from aqueous alteration (Le Guillou et al. 2015). While this might appear surprising given than CM chondrites are generally considered more altered than CR chondrites, our measurements are probing iron, whose abundance is higher in the matrix than in chondrules. From mass balance consideration, Fe-XANES measurements are thus likely sensitive to the alteration of the matrix. Using L-edge Fe XANES spectroscopy, Le Guillou et al. (2015) were able to decipher the valence state of iron in amorphous silicates and phyllosilicates of CR chondrites. Their work showed that iron is substantially oxidized in this phase ($Fe^{3+}$ / ($Fe_{tot}$) around 0.7). These values are higher than found in the present work ($Fe^{3+}$ / ($Fe_{tot}$) around 0.22), but our measurements were obtained on bulk CR. The measurements obtained here include the contributions of other phases like $Fe^{2+}$-bearing sulfides in the matrix, mixed $Fe^{2+}$, $Fe^{3+}$ in magnetite as well as possible $Fe^{2+}$ phase in chondrules. With the exception of GRA 06100, Renazzo and LAP 04156 contain the lowest $Fe^{3+}$ proportion but also the lowest metal abundance. If the initial proportion of metal were identical in CR chondrites, this points to different oxidation conditions that produced different amount of $Fe^{2+}$ and $Fe^{3+}$ than other CRs.

These results confirm that iron in the matrix of CR chondrites is significantly oxidized, and water was likely the oxidative agent (Le Guillou et al. 2015). The fact that CR chondrites matrix contain an significant fraction of $H_2O$ and –OH bearing phases is now established (Alexander et al. 2013; Bonal et al. 2013; Garenne et al. 2014; Le Guillou et al. 2015, Le Guillou and Brearley 2014). Even in MET 00426, described as the most primitive CR chondrite in our dataset (the leastaqueously altered) (Abreu and Brearley 2010; Le Guillou et al. 2015), the iron is significantly oxidized (Fig. 4). The pristine feature of MET 00426 matrix is debated (Bonal et al. 2013) and could have preserved those unstable phases (Scott and Krot 2005). However the two techniques (TEM and IR) are probing the samples at different scales. More generally the origin of meteorite matrices is debated, but a number of arguments seem to point toward a formation as disequilibrium condensates at least for the most primitive samples (Brearley 1993; Scott and Krot 2005, Abreu and Brearley 2010). It remains to be understood whether during the matrix formation process significant amount of $Fe^{3+}$ may have been incorporated into the amorphous phases, together with water. Formation of hydrated amorphous silicates in the disk does not appear to be an efficient mechanism (Prinn and Fegley 1989) to explain $Fe^{3+}$ phases in carbonaceous chondrites. The fact that GEMS grains, that have mineralogical affinities with the amorphous matrices of primitive carbonaceous chondrites are depleted in $Fe^{3+}$ (Keller and Messenger 2012; Hopp and Vollmer 2018; Vollmer et al. 2018), argues again hydration in the disk. Therefore, the hydrated and oxidized amorphous matrix observed here in CR chondrites likely represents the very first step of aqueous alteration on the parent body ( Le Guillou and Brearley 2014; Brearley 2014; Le Guillou et al. 2015).

## 5 Conclusion

The present study evaluates the bulk iron oxidation as well as abundance of $Fe^0$, $Fe^{2+}$ and $Fe^{3+}$ in CM, CR, CI and CV chondrites, based on XANES spectroscopy.

-A general trend between the iron oxidation index and the hydrogen content is observed for the different groups of carbonaceous chondrites. Fe is on average more oxidized in the order of CRs < CMs < CIs

-It appears that, after the first aqueous alteration stage, extensive aqueous alteration reduces the iron oxidation state of CM chondrites. This is most likely due to the dissolution and recrystallization of cronstedtite to serpentine.

-Based on the iron oxidation, it is not possible to distinguish CM chondrites finds from falls. There is no evidence that terrestrial residence had an impact on the speciation of Fe for our set of CMs.

-The heated CMs are not clearly distinguishable from the other CMs in terms of their $Fe^{2+}$ vs $Fe^{3+}$ relative abundance, but they generally contain a higher proportion of metal. There is also redox memory of the aqueous alteration process.

-The CR chondrites present a wide range of oxidation of iron (this could be compared to CM chondrites) controlled by the extent of aqueous alteration. $Fe^{3+}$ is present even in the least-altered CR chondrites.

- The iron oxidation index of CV chondrites is not consistent with the reduced/oxidized sub-classification. Rather, the iron oxidation of CVs seems strongly related to their thermal alteration.

**Acknowledgment**—This study was supported through funding from the H2020 European Research Council (ERC) (SOLARYS ERC-CoG2017_771691).

**Tables**

Table 1: Quantification of the relative atomic proportion of the iron oxidation state for CM, CR, CI, and CV chondrites with the iron oxidation index calculated (IOI). The heated CMs as been classified by Alexander et al. (2013)

**Figures:**

Figure 1: a) Normalized XANES spectra of selected carbonaceous chondrites with enlarged pre-edge feature. b) XANES spectra references for $Fe^0$, $Fe^{2+}$, and $Fe^{3+}$ oxidation state.

Figure 2: Normalized XANES spectra of CR chondrites with inset expanded diagram of the pre-edge feature.

Figure 3: Normalized XANES spectra of CM chondrites with inset expanded diagram of the pre-edge feature.

Figure 4: a) Average of iron oxidation index for the different groups of carbonaceous chondrites. b) Diagram of $Fe^{3+}/Fe^{2+}$ in comparison to $Fe^0/Fe_{tot}$ for the different groups of carbonaceous chondrites

Figure 5: Comparison between the average oxidation factor of iron bulk H content (from Alexander et al. (2012)) for the different groups of carbonaceous chondrites.

Figure 6: Comparison between the average oxidation factor of iron with the 3-µm water band from Beck et al. (2014) for the different groups of carbonaceous chondrites.

Figure 7: Comparison between the average oxidation factor of iron with the bulk D/H ratio from Alexander et al. (2012) for the different groups of carbonaceous chondrites.

a)

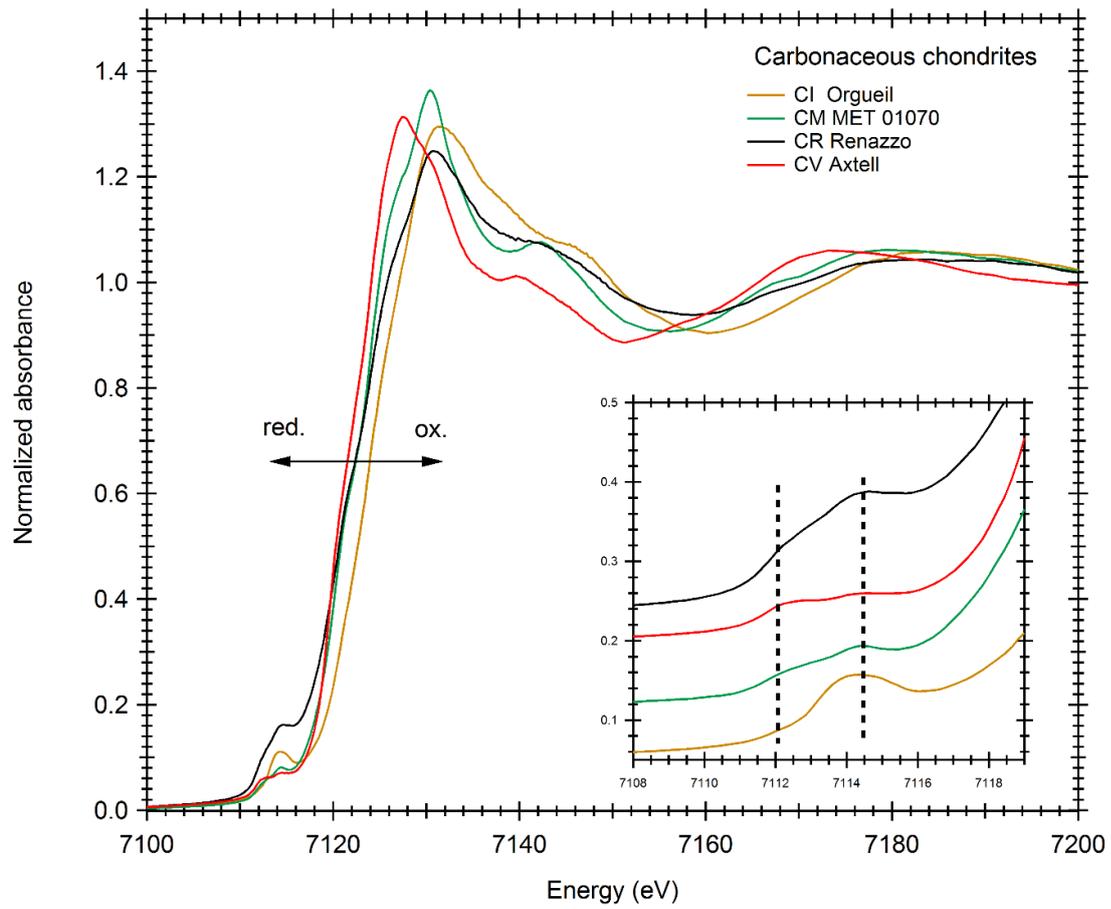

b)

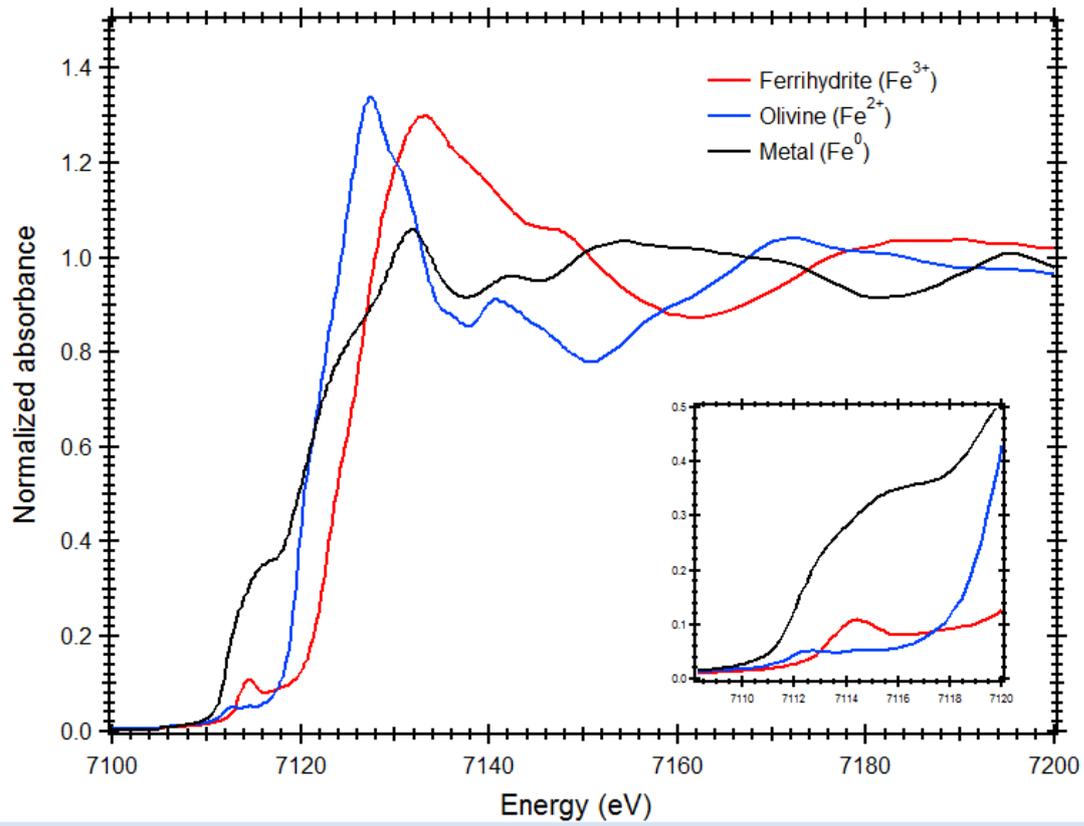

FIGURE 1

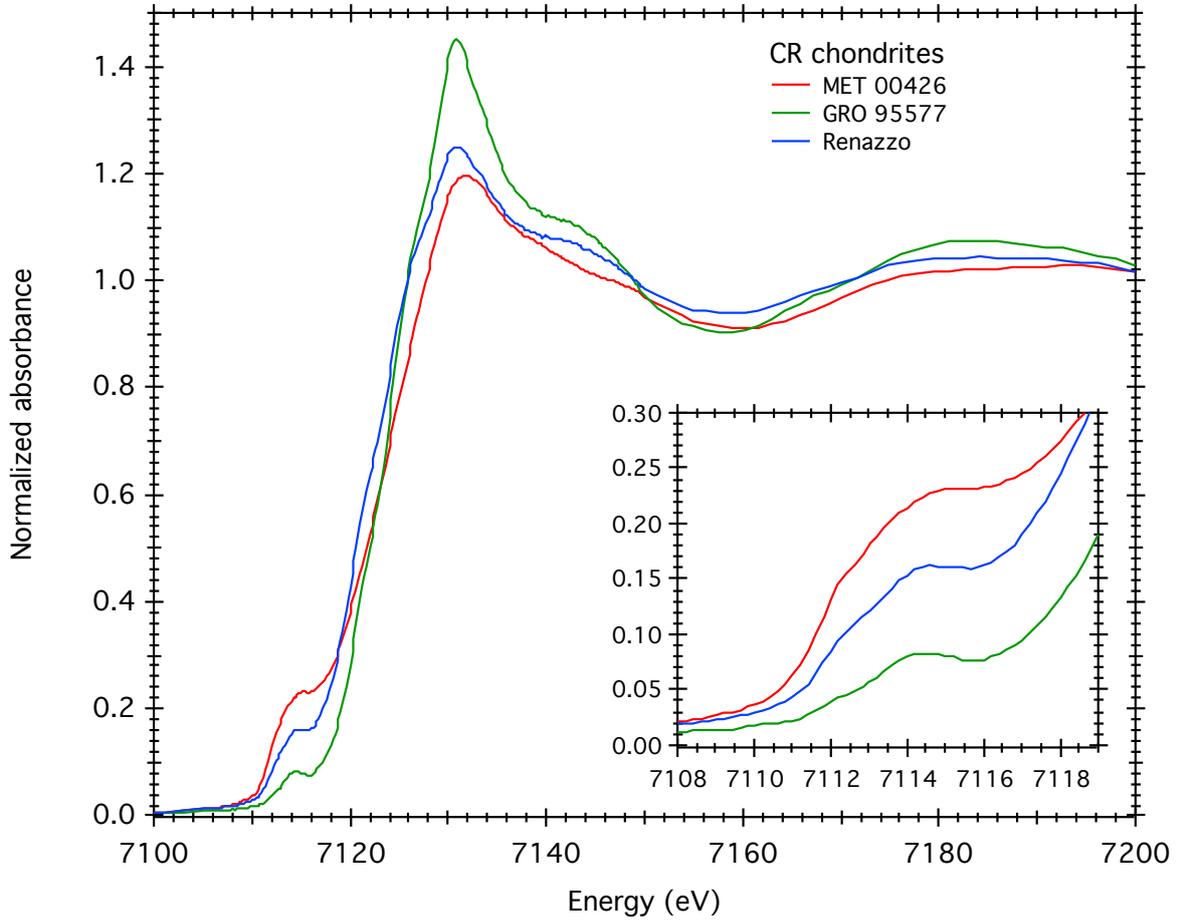

FIGURE 2

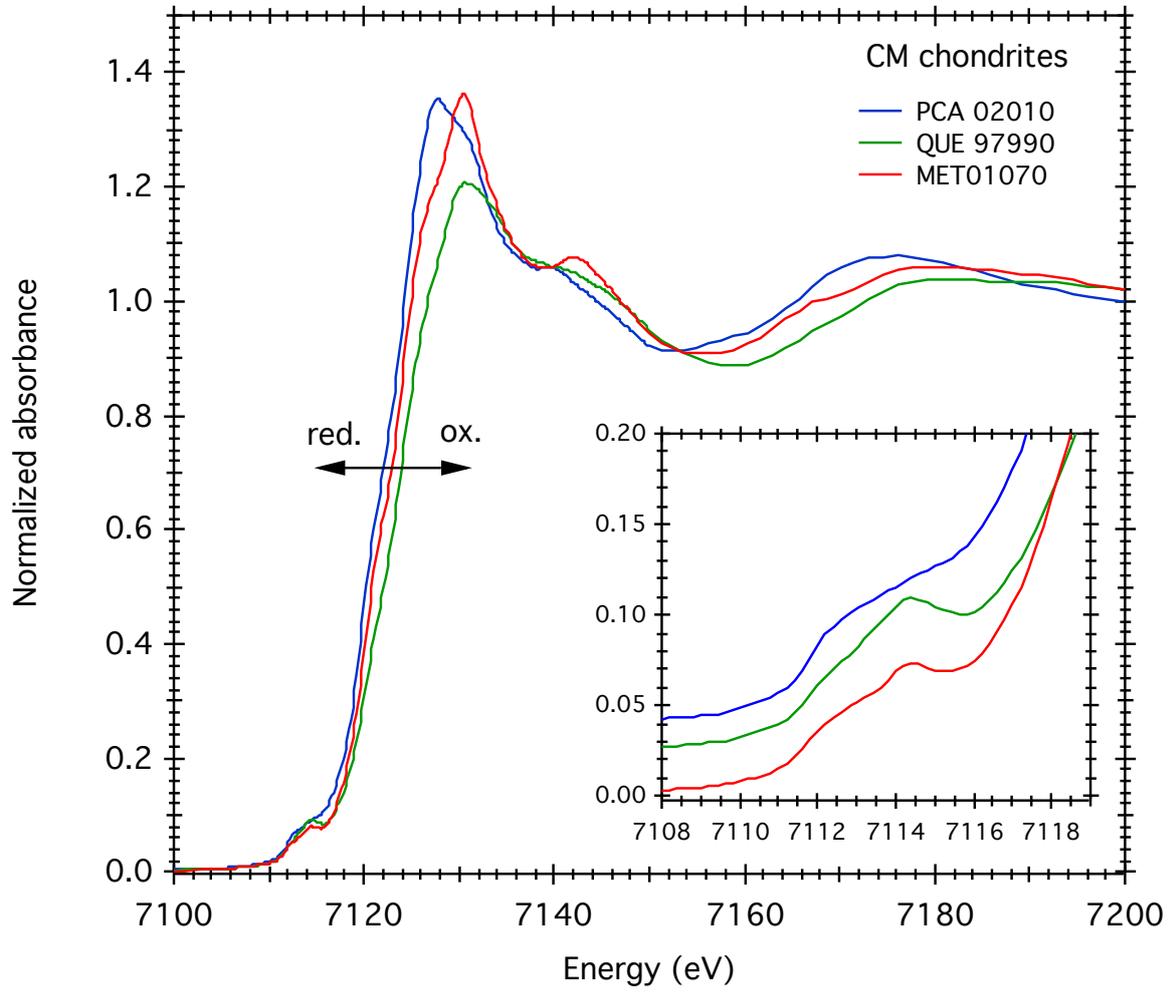

FIGURE 3

a)

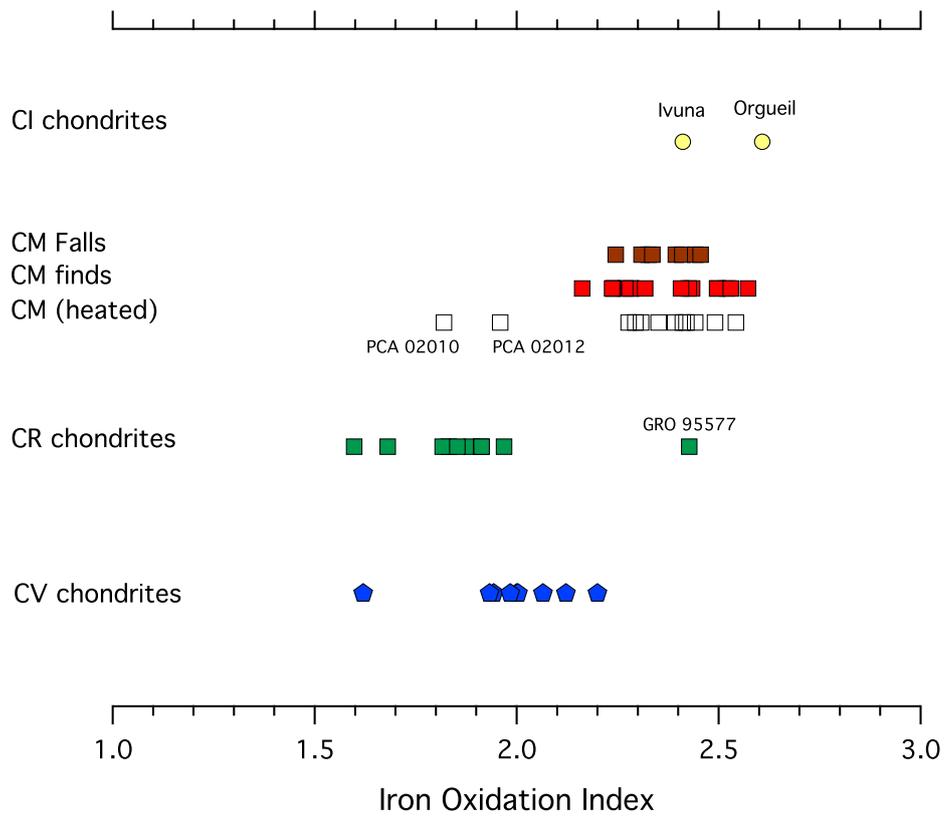

b)

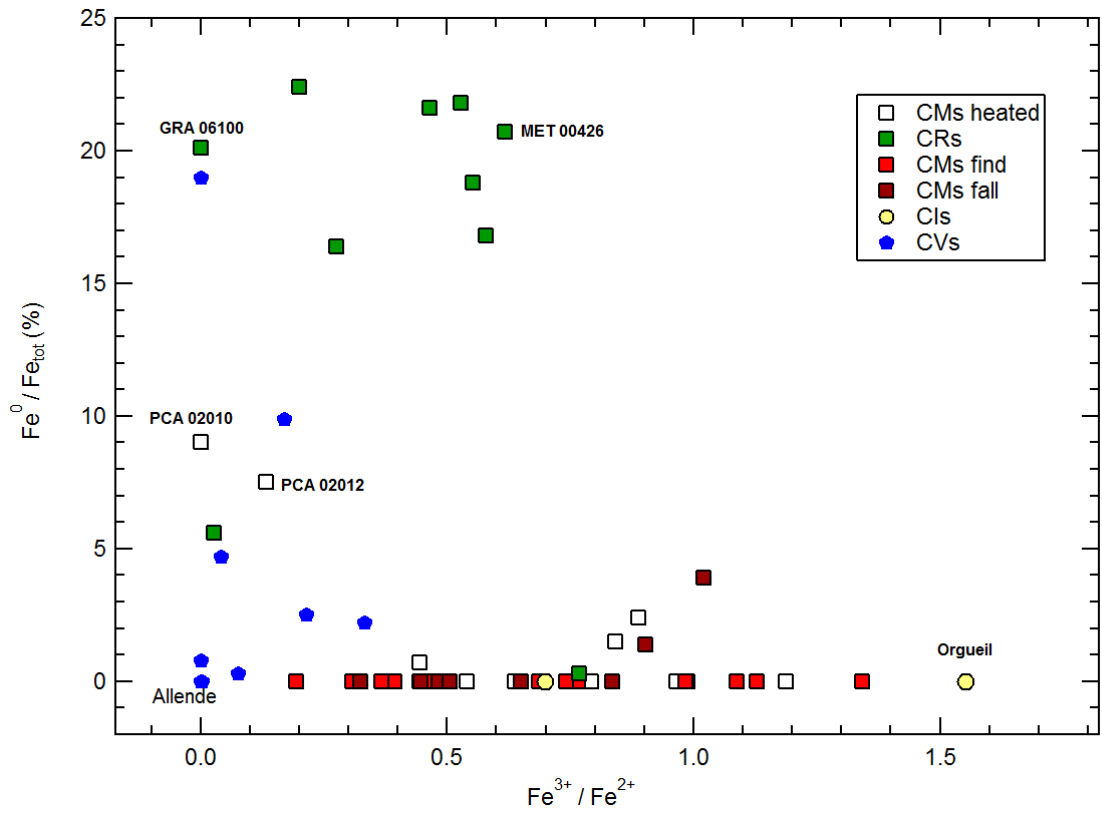

FIGURE 4

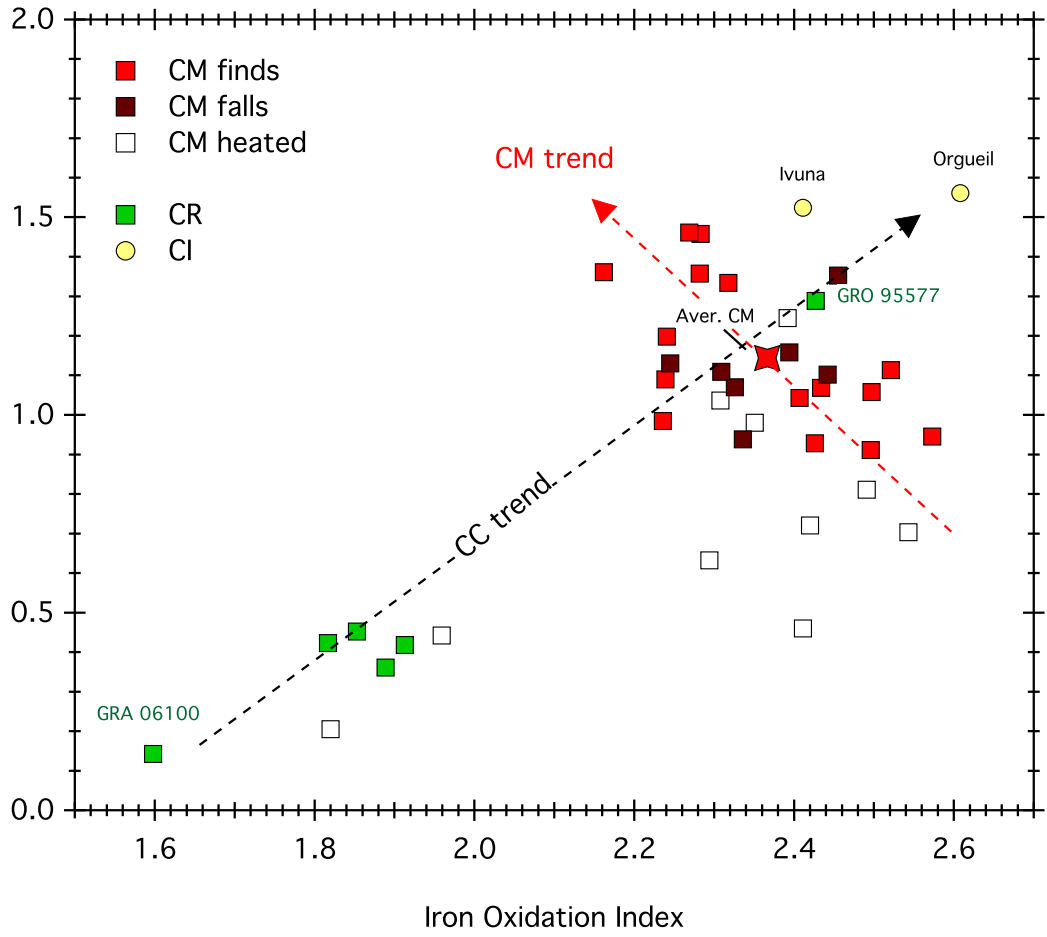

FIGURE 5

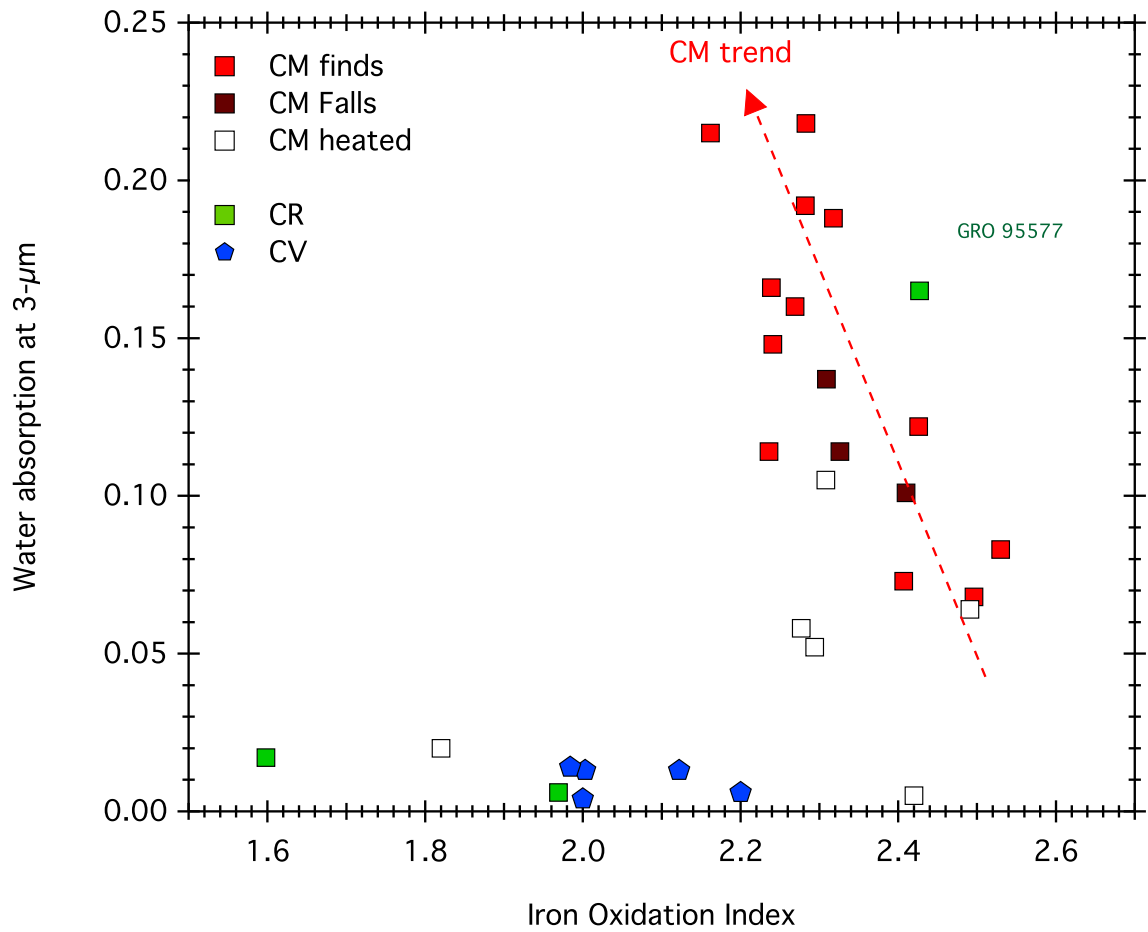

FIGURE 6

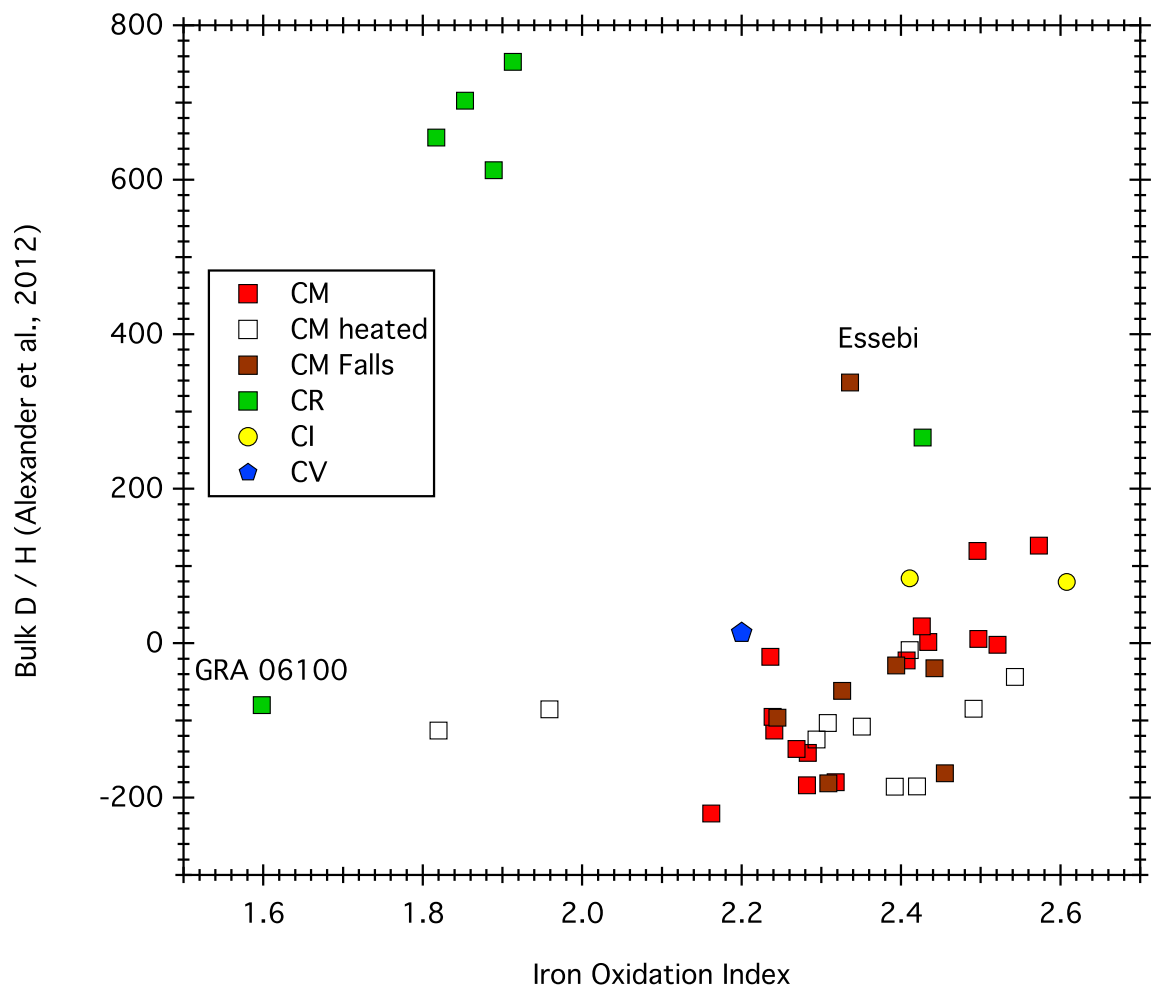

FIGURE 7